\documentclass[12pt]{article}

\usepackage{ifpdf}

\usepackage[utf8]{inputenc}
\usepackage{authblk}
\usepackage{subcaption}
\usepackage[onehalfspacing]{setspace}
\usepackage[margin= 1in]{geometry}

\usepackage{amsmath}
\usepackage{amssymb}
\usepackage{graphicx}

\usepackage{natbib}

\usepackage{float} 
\usepackage{multirow}
\usepackage{booktabs}

\title{Comparative study of training intensity in neurofeedback} 

\author[1]{Inês Esteves \thanks{Correspondence to: Inês Esteves, Instituto Superior Técnico, Avenida Rovisco Pais 1, 1049-001 Lisboa, Portugal; ines.esteves@tecnico.ulisboa.pt}}
\author[2]{Wenya Nan}
\author[1]{Cristiana Alves}
\author[1]{Alexandre Calapez}
\author[1,3]{Fernando Melício}
\author[1,4]{Agostinho Rosa}

\affil[1]{Evolutionary Systems and Biomedical Engineering Lab, Institute for Systems and Robotics, Lisbon, Portugal}
\affil[2]{Department of Psychology, Shanghai Normal University, Shanghai, China}
\affil[3]{ISEL – Instituto Superior de Engenharia de Lisboa, Instituto Politécnico de Lisboa, Lisbon, Portugal}
\affil[4]{Department of Bioengineering, Instituto Superior Técnico, University of Lisbon, Lisbon, Portugal}

\date{}

\begin{document}
\maketitle

\let\thefootnote\relax\footnotetext{This is a preprint version.}

\begin{abstract}
%\section*{}
Neurofeedback has proved to be useful in many instances. This technique is often used to address both medical issues and performance improvement. Despite the wide range of applications, no consensus has been reached about the optimal training schedule. 

In this work, a practical experiment was conducted aiming to compare the effects of intensive and sparse training modalities of enhancement of the individual upper alpha band at Fz, for working memory improvement, in 19 healthy subjects. The INTENSIVE group was submitted to 4 sessions of 37.5 minutes each, during consecutive days, while the SPARSE group performed 15 sessions of 25 minutes along approximately 2 months. 

The intensive modality proved to be significantly more effective in increasing the target frequency at Fz within session. However, no significant differences were found across sessions neither regarding cognitive improvements. The distinction between learners and non-learners led to significant results for the two groups across sessions, but only the learners of the INTENSIVE group show significant changes within session. Besides, for this subsample the effects in the INTENSIVE group are significantly stronger both within and across sessions.

The results suggest that longer sessions with shorter trials and more breaks could be advantageous when training healthy subjects. Although further investigation is needed to establish if one modality outperforms the other, better outcomes might be obtained if each session consists in many short trials (in this case 30 seconds) and is performed on consecutive days.

\small{ \section*{Keywords:}Neurofeedback, Training intensity, Individual upper alpha band, Working memory, EEG}

\end{abstract}

\section{Introduction}

Neurofeedback (NF) relies on the voluntary modulation of brain activity: brain signals are extracted and evaluated in real time, and presented back to the individual in the form of an auditory or visual feedback \citep{kober2017}. Since it does not depend on an external electrical or magnetic stimulus, it is expected to act without causing brain dependency \citep{niv2013}. Therefore, it reveals a great potential, as a complementary or alternative therapy, to deal with medical conditions when the conventional treatments are not successful or induce negative side effects \citep{mtacooperativegroup2004, arnold2013}. 
For instance, NF has been used to target attention deficit hyperactivity disorder \citep{meisel2013, mayer2016, strehl2017}, which is widely treated with stimulant medication \citep{strehl2006} that may cause problems such as growth suppression, impaired sleep and irritability \citep{nationalcollaboratingcentreformentalhealthlondon2009, mtacooperativegroup2004}. Another example is the current pharmacological treatment for schizophrenia which is considered insufficient and may cause the limitation and deterioration of functions when used for a long period \citep{nan2017}. As a consequence, new treatment options are being pursued, and NF has already shown positive behavioural outcomes when applied to this condition \citep{bolea2010, nan2017}. 

In addition to clinical applications, in recent years, research has spread to non-medical fields and newer protocols have been applied \citep{gruzelier2014c, mirifar2017}. However, despite the increasing widespread use of this technique, some aspects of NF methodology, such as the number of frequency bands, feedback modality, number of electrodes and training intensity are not standardized \citep{rogala2016}. Training intensity comprehends aspects such as the number of sessions, the duration of the subdivisions within one session and the spread of the training over time. For the same application, very different training intensity parameters have been used. For instance, for schizophrenia, one protocol consisting in a total of 13.5 hours within four consecutive days \citep{nan2017} and also another one composed of a total of 130 sessions of 1 hour twice per week, during 18 months, \citep{bolea2010} revealed positive results at the electrophysiological and behavioural levels. Although this example needs an adequate context since it refers to a complex pathology, it illustrates that in some cases much time and resources could be saved, as results could probably be achieved in a more efficient way, increasing also the motivation of the subjects. 

Furthermore, several authors attribute poor or unexpected results to insufficient training time \citep{dias2011, lecomte2011,nan2012}. Also, when making a distinction between learners and non-learners, it might even be the case that some individuals considered as non-learners, {i.e.} that do not respond to the protocol as expected, simply needed more sessions to consolidate results \citep{wang2016a}. Therefore, a deeper knowledge of training intensity would also allow for a more reliable evaluation of the results, by controlling the impact of choices regarding this matter.  

In this work, an experiment was performed with healthy subjects to test if contrasting training modalities (distinct with respect to the number of sessions, frequency and duration of each session) produce significant differences. These are quantified regarding electrophysiological changes at the cortical level and performance during a target task. The training aims to increase the individual upper alpha amplitude and, consequently, improve working memory. As there is no optimal protocol regarding intensity, this investigates if the learning process may occur within a short period. This is expected to be more easily accepted by participants as it is lighter and takes less effort regarding scheduling. 

To our knowledge, this is the first study that directly compares the effects of two distinct NF training intensity modalities. Although the option of studying several variables (number of sessions, frequency and duration of each session) at the same time may prevent the isolation of a single effect, it allows a preliminary investigation of the impact of distinct intensity factors. The examination of only one intensity parameter would rely on the assumption of a fixed model, dependent on a specific variable, and would reduce the chances of finding any effects, as there are many possible sources of impact on training. Accordingly, we selected two protocols as distinct as possible which would still be feasible in terms of total duration and availability of the participants.  

\section{Methods}

 \subsection{Participants}
A total of 19 healthy subjects participated in this study. Participants were allocated to two different groups: INTENSIVE (n = 9) and SPARSE (n = 10). It was not possible to randomly assign participants to their group: since the training load differed significantly between groups, time constraints had to be taken into account and thus the choice was made according to the participants' requirements. There was one drop-out in the SPARSE group after 4 sessions, due to incompatibilities with personal schedule. Considering the final sample, the INTENSIVE group consisted in 2 males and 7 females (age: $23.44 \pm 2.41$ range: $22$-$30$) while the SPARSE group consisted in 6 males and 3 females (age: $27.67 \pm 9.81$, range: $22$-$46$). The Wilcoxon rank sum test showed no significant difference between groups regarding age ($U = 38.5,\ p = 0.857$). 

This study was carried out in accordance with the recommendations and guidelines of the Ethics Committee of the CHLN and CAML with written informed consent from all subjects. All subjects gave written informed consent in accordance with the Declaration of Helsinki. The protocol was approved by the Ethics Committee of the CHLN and CAML. Participants were all volunteers and no monetary reward was given for their cooperation with this study. 
	
	\subsection{Design}
	The differences between INTENSIVE and SPARSE groups rely only on training intensity. INTENSIVE group performed 4 sessions, of approximately 2 hours each, in 4 consecutive days while SPARSE group performed 15 sessions, around 1h30 each, spread along approximately 8 weeks, with 2-4 sessions per week. A follow-up assessment was carried approximately one month after they completed the training. Both groups were submitted to NF training (upregulation of the UA band with feedback), transfer trials (upregulation of upper alpha (UA) without feedback) and cognitive tests (tests to assess working memory performance). Resting baselines were preceded by a 2-minute relaxed state and assessed both Eyes Open (EO) and Eyes Closed (EC) conditions, with recording of two alternating epochs of 1 minute for each, both at the beginning and at the end of each session. A questionnaire to assess general health state was answered in the first session (before training) and in the follow-up assessment. In every session, participants were also asked to fill a questionnaire in which they rated several parameters referring to their mental state during the session. 
	
\begin{figure}[h!]
	\begin{center}
		\includegraphics[width = 1\linewidth, keepaspectratio]{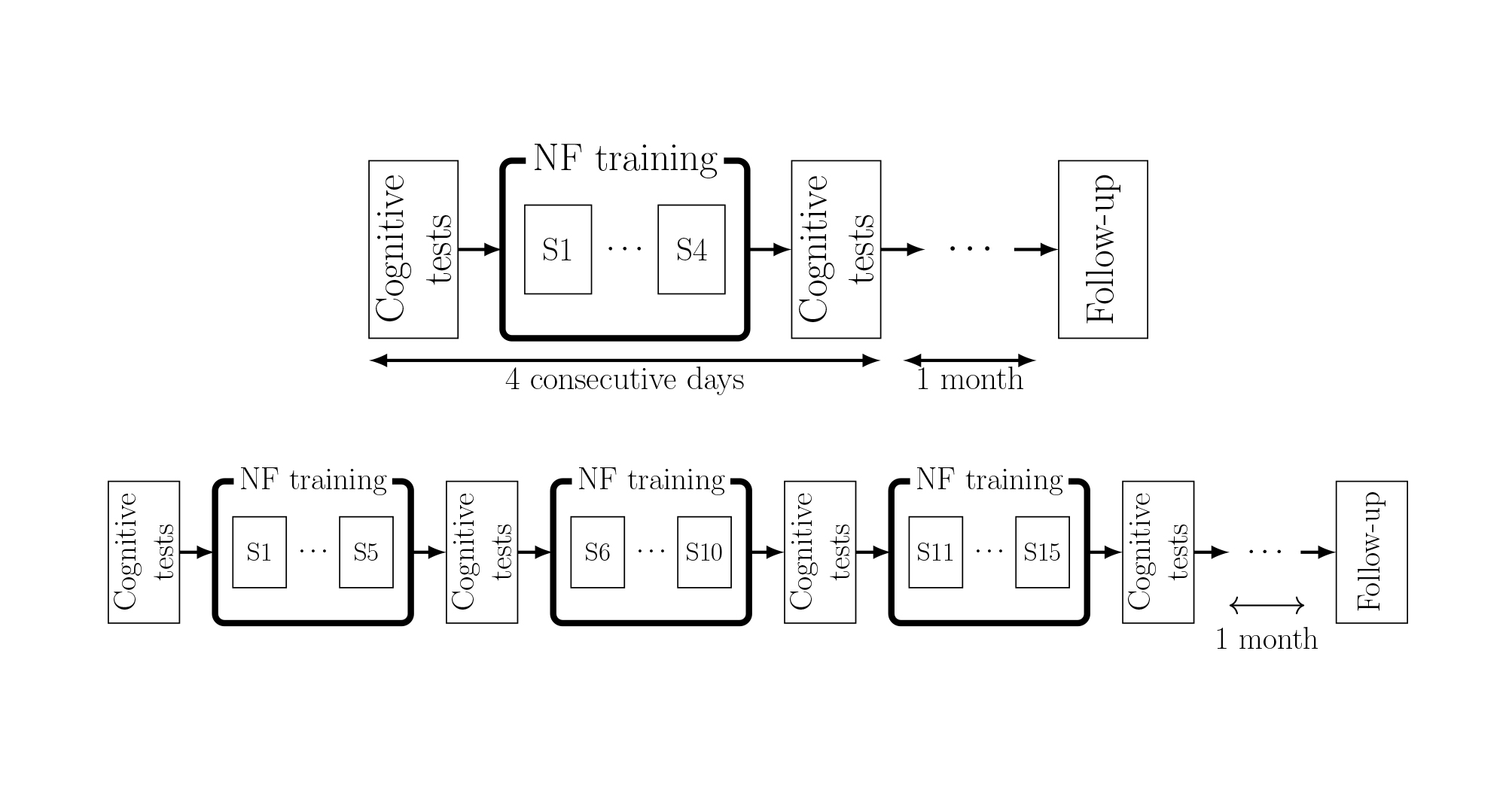}
	\end{center}
	\caption{Top: Training diagram for the INTENSIVE group; Bottom: Training diagram for the SPARSE group}
		\label{fig:training}
\end{figure}

	Diagrams illustrating the training workflow for the INTENSIVE and SPARSE groups are represented in Figure \ref{fig:training}, top and bottom, respectively. Although the cognitive tests were performed at the beginning or at the end of one session, they are represented outside the respective session. This is done to facilitate the understanding of the schematics, since these evaluations were not part of every session.
	
\begin{figure}[h!]
	\begin{center}
		\includegraphics[width = 1\linewidth, keepaspectratio]{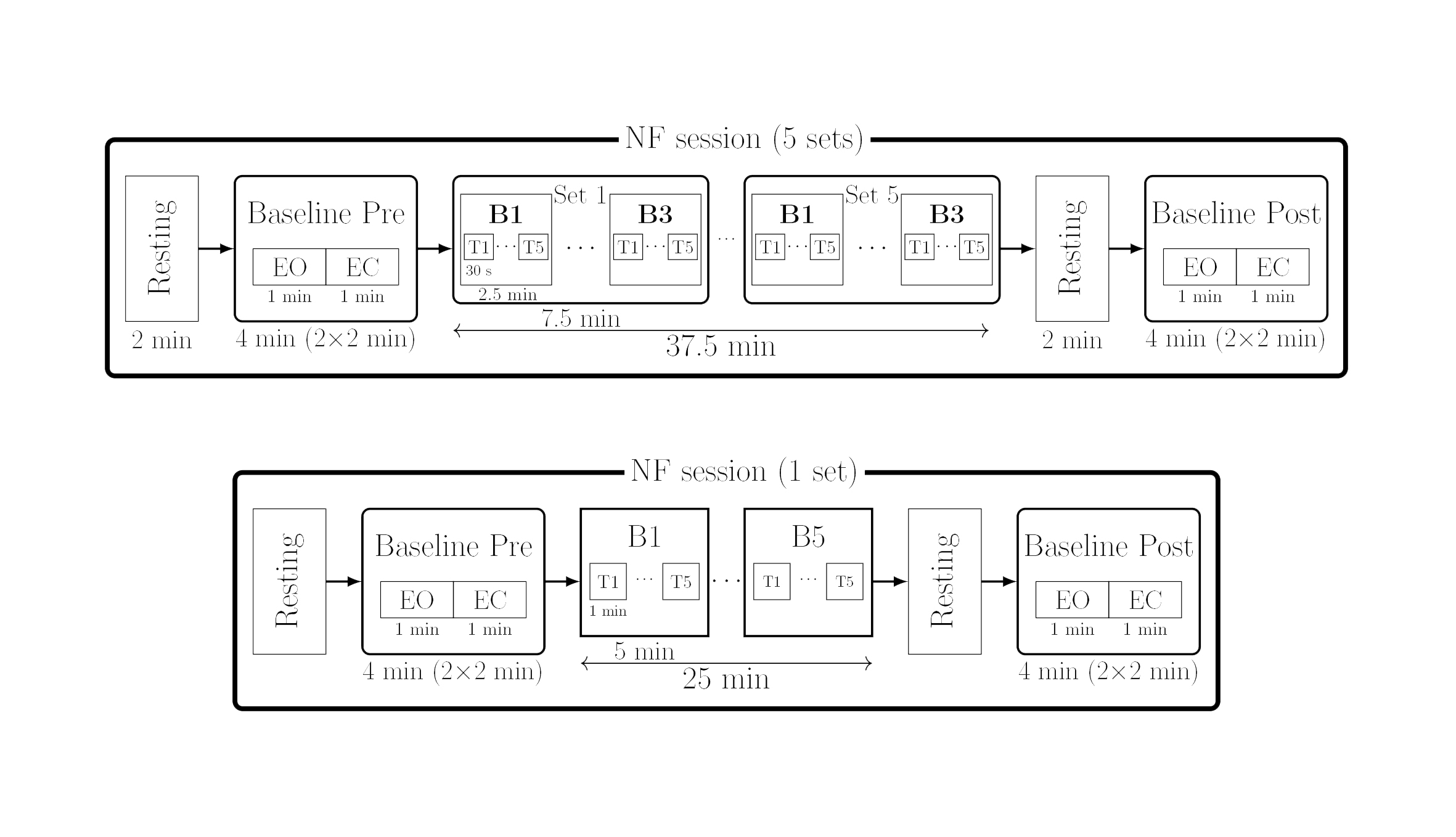}
	\end{center}
	\caption{Top: Session diagram for the INTENSIVE group; Bottom: Session diagram for the SPARSE group}
		\label{fig:session}
\end{figure}

	Participants in the INTENSIVE group underwent sessions with 5 sets of 3 blocks each. The blocks were composed by 5 trials of 30 seconds, with 5-seconds intervals between them. Transfer trials were introduced from session 2 to 4, during which no feedback was provided. A transfer trial for adaptation, with a duration of 30 seconds, was introduced at the end of the 3rd set of blocks in the 2nd session. The following transfer periods consisted in two trials of 30 seconds and were performed after the 5th set of the 2nd session, after the 3rd and 4th sets of the 3rd session and after the 3rd and 4th sets of the 4th session. The total NF time in each session was 37.5 minutes (not taking into account transfer trials and intervals between trials). Cognitive tests were performed after the pre-training baselines in the first session and after the post-training baselines in the last session. The follow-up assessment was similar to the first training session, with the exception that the training period consisted in a single training block, followed by a 30-second transfer trial. Figure \ref{fig:session} (Top) illustrates the plan for each session of the INTENSIVE group. Again, in order to facilitate the representation, intermittent assessments are not included in the schematic, which in this case correspond to the transfer trials, whose temporal organization within training varies among sessions.
	
	Participants in the SPARSE group underwent sessions with a single set of 5 blocks. Each block consisted in 5 trials of 60 seconds each, with 5 seconds of interval between them. Transfer trials were introduced only in the 11th session and lasted until the last session. In every session, after the 5th NF training block, a transfer block was performed, with 2 transfer trials of 1 minute separated by 5 seconds of interval. Cognitive tests were performed after the pre-training baselines in the first session and after the post-training baselines in the 5th, 10th and 15th sessions. As in the INTENSIVE group, the follow-up assessment also resembled the first session, differing only in the existence of a single training block which was followed by a 1-minute transfer trial. Figure \ref{fig:session} (Bottom) illustrates the plan for each session of the SPARSE group, transfer trials are also not included.

 	\subsection{Signal acquisition}
	The acquisitions were carried out using \textit{Somnium} software \citep{rodrigues2010}, in a room provided by the Evolutionary Systems and Biomedical Engineering Lab (LaSEEB), a research lab of Institute for Systems and Robotics (ISR), at Instituto Superior Técnico (IST), University of Lisbon. Electrodes were placed according to the International 10-20 System, using the left and right mastoids as references for common mode rejection and the middle of the forehead as ground. Relevant signal was recorded, with a sampling frequency of 250 Hz, from 20 electrodes: Fz, Fp1, F7, F3, T3, C3, T5, P3, O1, Cz, Pz, Oz, Fp2, F8, F4, T4, C4, T6, P4 and O2, and amplified by the EEG amplifier Vertex 823 (produced by Meditron Electromedicina Ltda, São Paulo, Brazil), with an analogical band-pass filter between 0.1 and 70 Hz. The impedance was measured during the process, guaranteeing that all electrodes had an impedance below $10\ k\Omega$. Subjects were asked to remain as still as possible and also to avoid excessive blinking and abrupt movements.
	
	\subsection{Protocol}
	Although the main focus of the present work is the study of training intensity effects, it also tries to evaluate the outcomes of distinct training intensity modalities regarding the enhancement of working memory (so that healthy participants felt more engaged with the tasks). For that purpose, training consisted in the enhancement of the individual upper alpha band at Fz, as both the upper alpha band and the frontal area are associated with memory functions \citep{zola-morgan1993}. The evidence of a link between increased upper alpha band activity and good working memory performance \citep{klimesch1999, escolano2011, hsueh2016} lead to the decision of training individuals to increase the activity in this band. The use of the Individual Alpha Band (IAB) was based on \cite{klimesch1999a} that invokes a large variability of the boundaries of this band among subjects to justify the need for a personalized definition.    

		\subsubsection{Individual upper alpha measurement}
The first baseline measurements were used to define the IAB of each individual. The power spectrum density is estimated using the Welch's method \citep{welch1967}, with an overlap of 10\% and a segment length of 5 seconds, in \textit{Somnium}. The crossings between EO and EC spectra provide the frequency boundaries Lower Transition Frequency (LTF) and Higher Transition Frequency (HTF).  The UA band is defined as the frequency range between the individual peak alpha frequency, which was considered to be equivalent to Individual Alpha Frequency (IAF), and the HTF. 
	
		\subsubsection{Neurofeedback training}  
		
		\paragraph{Feedback}\hfill\\
		The EEG training platform integrated in the \textit{Somnium} software was used to perform NF training. The feedback is established according to the achievement of two goals and makes use of two three-dimensional objects against a grey background, a sphere and a cube, that change according to previously defined settings and react to subject's EEG in real-time \citep{rodrigues2010}. \textit{Goal 1} depends on the relation between the EEG measurement that is being evaluated and a pre-defined threshold value. In this case, since the interest is to enhance the upper alpha band, it is considered that \textit{Goal 1} is achieved every time that the relative amplitude of this band is above the threshold. This makes the sphere change from white to purple and increase in size and number of sides. \textit{Goal 2} is achieved everytime that \textit{Goal 1} is sustained for more than 2 seconds, making the cube go up until \textit{Goal 2} is no longer reached. Using $X(k)$ to denote the frequency amplitude spectrum computed using the Fast Fourier Transform with a sliding window of 2 seconds and shifts every 0.125 seconds, the relative amplitude as presented in \cite{wan2014} is given by:
		
\begin{equation}
Relative\ Amplitude = \dfrac{Band\ Amplitude}{EEG\ Amplitude}
\end{equation}
		
		In this case, for the upper alpha relative amplitude:
		
\begin{equation}
 UA\ Relative\  Amplitude = 
		\dfrac{
			\dfrac{\sum_{k=IAF/\Delta f}^{HTF/\Delta f} X(k)}{HTF-IAF}
			}
			{
			\dfrac{\sum_{k=3.99/\Delta f}^{30/\Delta f} X(k)}{30-3.99} 
			}
\end{equation}
		The frequency resolution is represented by $\Delta f$. The EEG amplitude is measured considering the frequencies between 3.99 and 30 Hz (lower frequencies were discarded to mitigate the effects of blinking). 
The baseline amplitude values were defined as described in the previous subsection and maintained along the sessions for both groups, as we assume there should not exist significant intra-variability. 

All subjects started with a threshold value of 1, which was adjusted according to individual performance evaluated by the total amount of time during which \textit{Goal 1} was reached. If the percentage of time during which the feedback parameter was above threshold exceeded 60\%, the threshold was increased by 0.1. If this percentage was lower than 20\%, the threshold was decreased by 0.1. This was done in order to keep it challenging if the performance is considered good and, if the opposite happened, to allow the subject to find the most successful mental strategies without losing motivation along the process. The aim was to maximize the effect of the feedback. 
		    
		\paragraph{Training}\hfill\\
		Participants were informed about the existence of the two goals during NF training and told to focus primarily on the sphere, trying to make it  purple, and as round and big as possible. They should use a single mental strategy per block, so that the effects of that strategy could be isolated in order to rate its effectiveness. During the first blocks, participants were encouraged to try different strategies in order to understand which ones produced better results and then repeat them afterwards. They were allowed to make short breaks between blocks, if needed, while the practitioner defined the parameters for the following block and wrote down the mental strategy that was used. 
		
		In the INTENSIVE group, between the sets of blocks there was a larger break that also helped to check the average time spent above threshold and reset the NF threshold when necessary. In the SPARSE group, threshold resetting only occurred at the end of each session. 
		
		Transfer trials were performed to assess the ability to control cortical activity without feedback \citep{rockstroh1993, siniatchkin2000}.
		
		\paragraph{Mental strategies}\hfill\\
		Although the participants were not encouraged to use any specific strategies, some examples were provided when they asked for them. Since they were allowed to choose the more suitable strategies for them, the preferentially applied strategies varied a lot. The most successful strategy of each session (to which corresponded the block with highest UA amplitude) was collected for each individual. After gathering all the best strategies for every subject, 49 distinct strategies were found which were then grouped into six categories: ``feedback'', related with feedback display and the screen; ``imagination'', for fantasizing with unreal episodes; ``memories'', for recalling past experiences; ``mental'', when performing tasks that involved mental effort; ``motor'', for thinking about performing physical activities; ``relaxation'', attempting to relax the body and mind (for example, with breathing exercises). For both groups, ``relaxation'' was the preferred category (INTENSIVE: 38.89\% and SPARSE: 34.07\%). When counting the total number of different best strategies there was not a distinction between repeated use by the same participant or single use by several participants. 
		
		 \subsection{Measures}
  
 			\subsubsection{Working-memory performance}
 		 So that the effects of NF training in working memory performance could be analysed, allowing also a comparison between the two training intensity modalities, different cognitive tests were carried out: Digit Span (Forward and Backward) and N-Back (Numbers and Letters). During the Digit Span test \citep{to2016} participants had to recall several sequences, with increasing length in each trial, from 3 to 13 digits. Subjects were asked to introduce the digits in the order by which they appeared (forward digit span) and by the reverse order (reverse digit span). When the test finished, two performance measures were provided: digit span, the maximum amount of digits of a single sequence that one was able to remember, and the score, the percentage of correct correspondences between the real sequence and the one introduced by the subject.
 		
 		During N-Back test \citep{kirchner1958} the individuals were required to monitor a series of numbers/letters and identify if the current one was the same that was presented n trials before, with n = 3. Twenty-three trials were performed, which resulted in 20 answers. Each digit was shown for a maximum of 3 seconds, during which the participant could answer, and there was an interval of 2 seconds between trials. 

 		 \subsubsection{Questionnaires}
In other to keep record of how factors such as concentration, motivation, sleepiness and stress affect training, a questionnaire to assess these factors was performed. For this purpose, a rating scale was used to evaluate the frequency of the four mentioned states/sensations during training: 1 - never, 2 - rarely, 3 - sometimes, 4 - frequently, 5 - always. 

Furthermore, the 36-Item Short Form Survey (SF-36) questionnaire, which has been validated for the Portuguese population, was performed to assess different domains of health state and quality of life \citep{ferreira2012, randhealthcorporation2017}. It was performed in order to ensure that the participants did not have abnormal health conditions and at the same time compare their health state before and after the NF training.
 		 
		\subsection{Data pre-processing and extraction}
	The pre-processing and extraction of information from the raw EEG was performed using \textit{Somnium} tools. 
	It was necessary to remove some recorded periods which contained artifacts that would mislead the analysis. With the exception of eye movements, the artifact removal was performed manually, for all baseline measurements of each participant. 	
	Individual reports were transferred from \textit{Somnium} to Excel files and extracted with MATLAB software (version R2015b). The studied relative amplitudes correspond to the frequency bands presented in Table \ref{tab:frequency_bands}.	
		 
		\subsection{Data treatment}
		The treatment of the data for this work, including the visualization of plots as well as statistical analysis, was performed using the MATLAB software. The topographic distributions were generated using Fieldtrip \citep{oostenveld2011}, an open-source MATLAB toolbox.  
				
			\subsubsection{Evaluation of training performance}
		Since the participants had different baseline amplitude values and the electrode impedance varied from session to session, as the only restriction imposed was that values were under 10 k$\Omega$, relative amplitude values for each individual session were normalized. The normalization was enforced through division by the corresponding pre-training baseline with EO. Therefore, for simplification, normalized relative amplitudes will be hereafter referred as amplitudes only. 
		
		There is not a standard approach for the evaluation of training performance. Several authors relied on clusters of sessions or representative sessions, such as the first and the last, to extract conclusions about training effectiveness \citep{strehl2006a, leins2007, staufenbiel2014, hsueh2016}. On the other hand, others use measures that consider the progress along the whole training \citep{escolano2011, nan2012}. In this case, in order to evaluate the evolution of the amplitude of frequency bands along time, both within session and across sessions, distinct indexes were used to capture different aspects. We derived two types of measures, one that explores the overall trend and another to take into account the small variations that may occur. For $S$ sessions and $B$ blocks, denoting the $i-th$ session by $s_i$ and the $j-th$ block of that session by $b_{j,i}$: 

\begin{itemize}
	\item Within Session
	\begin{itemize}
		\item[--] W1: this measure is based on \cite{wan2014} and aims to quantify the mean change within session across all training sessions:

\begin{equation}
 W1 = \dfrac{\sum\limits_{i = 1}^{S}\sum\limits_{j = 2}^{B} (b_{j,i} - b_{1,i})}{S \times (B-1)}
\end{equation}

		\item[--] W2: is the slope of the linear regression that describes the evolution of the amplitude of a certain frequency band along blocks, averaged across sessions. Considering $ \textbf{y} = m_i\textbf{x} + b_i $ as the trendline for $s_i$:
	\begin{equation}
		W2 = m_{average} = \dfrac{\sum_{i=1}^{S} m_i}{S}
	\end{equation}
		in which $\textbf{y}$ corresponds to the amplitude, $m_i$ is the slope, $\textbf{x}$ is the block number and $b_i$ is the y-intercept, which will depend mostly on each subject's characteristics. 
	\end{itemize}
	\item Across Sessions
	\begin{itemize}%[leftmargin=*]
	\item[--] A1: the amplitude change of the last two sessions relative to the first two sessions. 
	
\begin{equation}
A1 = \dfrac {(s_{S-1} + s_S) - (s_1 + s_2)}{(s_1 + s_2)}
\end{equation}
	
	\item[--] A2: corresponds to the slope of the linear trendline computed for the evolution of frequency amplitude along sessions, therefore taking into account the variations that may occur in between. 
	Taking $ \textbf{y} = m\textbf{x} + b $, A2 corresponds to $m$
\begin{equation} 
A2 = m
\end{equation}	
	with $y$ corresponding to the relative amplitude of a certain frequency range, $x$ representing the session number and $b$ standing for the y-intercept, which, as in the previous case, will rely on each subjects' intrinsic characteristics.
	\end{itemize}	
\end{itemize}
 
The distinction between learners and non-learners relies only on the training performance for the target frequency band, in this case the UA band. Similarly to what was mentioned for the evaluation of training performance, there is also not a consensus regarding how to distinguish learners from non-learners and distinct methods have been applied \citep{escolano2011, hsueh2016, mayer2016}. In the present work, A2 measure being positive (positive slope) was chosen as the criteria to classify a participant as learner, as it reflects the learning trend across time and thus summarizes the performance. Five learners were identified in the INTENSIVE group (56\%) and six in the SPARSE group (67\%).

\subsubsection{Evaluation of cognitive effects}
The results of Digit Span Test and the N-Back test were used to evaluate the effects on cognition, at the working-memory level. For the Digit Span Test, two indicators were chosen ``Span'' and ``Score''. Regarding the N-Back Test, the performance was evaluated based on the ``Number of Successes''. The results obtained at $t_i$ (initial session), $t_f$ (final session) and $t_{fu}$ (follow-up session) were used to evaluate the effects along time and examine if the training influenced them.

 		\subsection{Statistical analysis}
This experiment was performed with a small sample (total $N = 18$ participants in the final sample), and several results are analysed individually for each group ($N = 9$ for each group). Shapiro-Wilk test \citep{shapiro1965} was used in order to assess if variables under examination came from a normally distributed population. Since in several cases normality was not verified, an option was made to rely only on non-parametric tests.	 
		
To analyse the median of a group or the difference between two time points within one group, Wilcoxon signed rank test was used \citep{conover1981}. Right-tailed tests were employed when there was a prior hypothesis of an increase, as was the case for the UA amplitude within and across sessions as well as (post$-$pre) baseline measurements. For validating of the use of the individual upper alpha band, the boundaries IAF and HTF were not compared with zero, but with standard values 10 and 12 Hz, respectively \citep{klimesch2005}. 
		
In order to make comparisons between groups, the Wilcoxon rank sum test was used, mostly to test if the medians considering all the participants of each group differed significantly \citep{chernick2003}. A significance level of 5\% was considered.	

In the case of topographical analysis regarding the UA amplitude, since multiple comparisons were performed as there were 20 electrodes, the false discovery rate (FDR) was controlled with the Benjamini-Hochberg procedure \citep{benjamini1995}. Considering $m$ hypotheses $H_1, H_2, ..., H_m$ to which correspond the p-values $P_1, P_2,..., P_m$, the p-values are sorted in ascending order $P_{(1)} \leqslant P_{(2)} \leqslant ... \leqslant P_{(m)}$, with $P_{(i)}$ corresponding to $H_{(i)}$. For a given threshold $q$, one should find $k$, the largest $i$ for which:

 \begin{equation}
 	P_{(i)} \leqslant \dfrac{i}{m}  q 
 \end{equation}

and reject all the hypotheses $H_{(i)}$ with  $i = 1, ..., k$. The threshold $q$ was defined as 0.05.

\section{Results}

		\subsection{Individual alpha band}
		Concerning only the initial IAB, at  $t_i$, the results indicate that the groups did not differ significantly regarding LTF ($U = 39,\ p = 0.914$) and IAF ($U = 30,\ p = 0.372$). However, the difference for HTF values was considered significant ($U = 17.5,\ p = 0.041$), at $t_i$. 
		 
		 In order to assess the effect of training on the intra-variability of the individual alpha band, it was tested if the median of the change ($t_f$ - $t_i$) was different from zero, for each group individually. There was only evidence of a statistically significant difference between HTF measured at $t_i$ and $t_f$ in the SPARSE group ($W = 33,\ p = 0.039$). However, it is possible that this change was not caused by the training intensity modality itself, but rather by intrinsic small changes along time. These were more likely to appear in this group since the $t_i$ and $t_f$ are separated by approximately 2 months. At $t_f$, there were no significant differences between groups for any of the frequencies.

		Although a common fixed reference range for the upper alpha band is from 10 to 12 Hz \citep{klimesch2005}, this experiment used the individual upper alpha band, so that each subject had a personalized intervention. To examine if the individual range was similar to the reference one, it was tested if the IAF and HTF of both groups (as a whole) differed significantly from 10 and 12 Hz, respectively. At $t_i$,IAF was not considered significantly different from 10 Hz ($W = 128,\ p = 0.064$). While for HTF the difference from 12 Hz was considered significant ($W = 140,\ p = 0.003$). Group values at $t_i$ were LTF: $Mdn = 8.25,\ IQR = 7.9-8.8$, IAF: $Mdn = 10.4,\ IQR = 9.7-11.2$ and HTF: $Mdn = 12.9,\ IQR = 12.5-13.3$. 

		\subsection{Training}
		The analysis of NF training results focuses on the changes that occur both along the time course of the training experiment and within each session. The evolution of the relative amplitude of UA was examined to assess trainability while changes regarding other frequency bands are shown so that independence may be studied, as suggested by \citet{zoefel2011}. The topographical effects of training were also inspected in order to understand whether they are restrained to the training location, Fz, or if there was a spread to other areas. 
		     
		\subsubsection{Across sessions}
\begin{figure}[h!]
	\begin{center}
		\includegraphics[width = 1\linewidth, keepaspectratio]{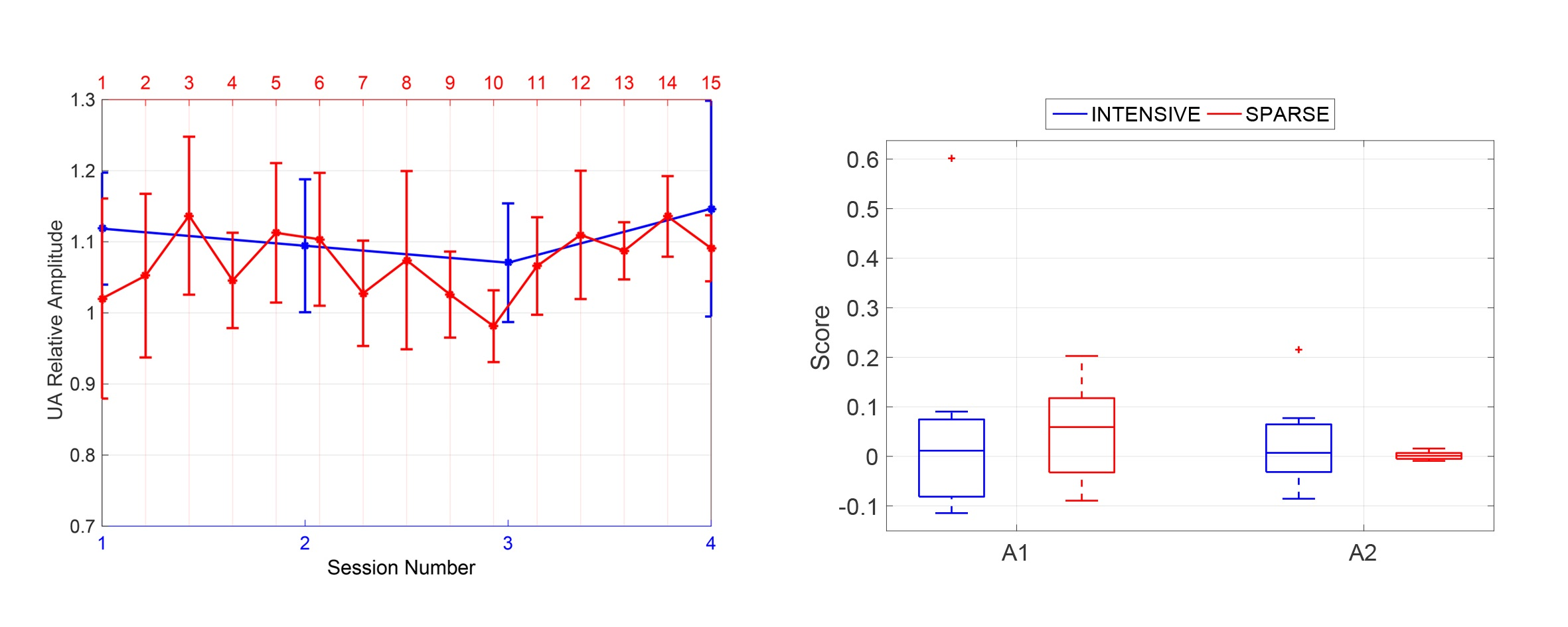}
	\end{center}
		\caption{Left: Evolution across sessions for the INTENSIVE (blue) and SPARSE (red) groups during neurofeedback at Fz; median values of UA amplitude; error bars show median absolute deviation. Right: Boxplot with distribution of A1 and A2 measures for the INTENSIVE (blue) and SPARSE (red) groups; the + represents outliers}
		\label{fig:nfacross_comparison}
\end{figure}

			\paragraph{Training location}\hfill\\
The evolution of the UA amplitude across sessions for both groups is depicted on Figure \ref{fig:nfacross_comparison}, with different scales, as the number of points is much smaller for the INTENSIVE group. For both groups there are amplitude fluctuations along time for the whole training period, although the difference between the initial and final values is larger for the SPARSE group. Accordingly, in Figure \ref{fig:nfacross_comparison}, the A1 measure shows a higher median for the SPARSE group and its distribution is shifted upwards when compared to the INTENSIVE group. For the A2 measure, both medians are close to zero and the distribution is highly concentrated for the SPARSE group, which indicates that this measure might have been a poor choice to discriminate between learners and non-learners since the differences are minimal. Overall, no significant effects across sessions were found for both A1 and A2, either within groups (Table \ref{tab:across_within_all}) or between them (Table \ref{tab:across_between_all}). 	

\begin{figure}[h!]
	\begin{center}
		\includegraphics[width = 1\linewidth, keepaspectratio]{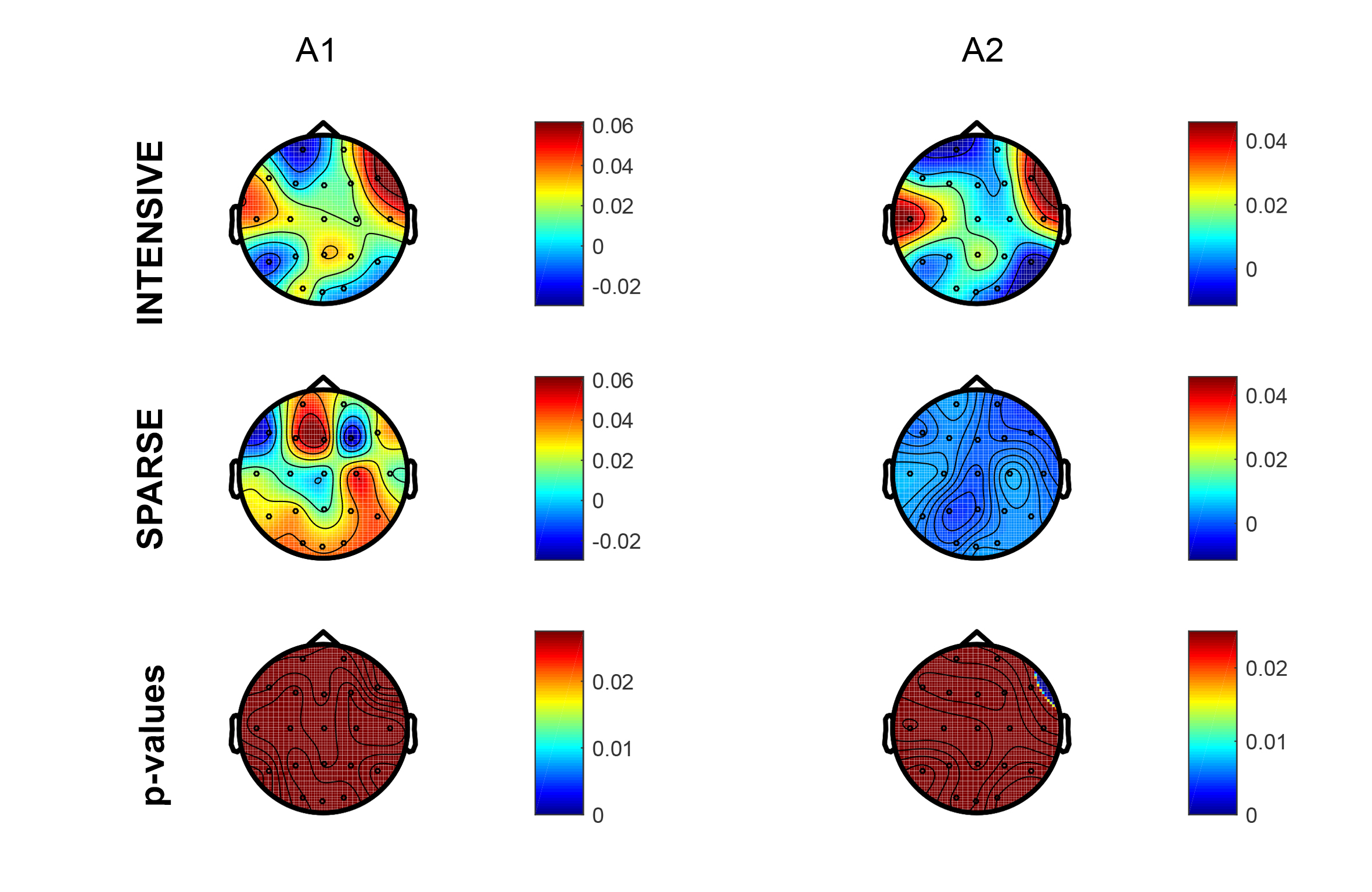}
		\end{center}
		\caption[Evolution across sessions]{Topographic distribution of median measures across sessions for the UA band}
		\label{fig:topoUA_A}
\end{figure}

			\paragraph{Topographic distribution}\hfill\\
Concerning the measures of performance across sessions, represented in Figure \ref{fig:topoUA_A}, the topographical results are according to what was previously noted from Figure \ref{fig:nfacross_comparison}, as there is an area of increased A1 around Fz for the SPARSE group. However, there are values of similar magnitude for the INTENSIVE group in the right frontotemporal and left temporal areas, which were not expected due to training. This pattern is also visible for the A2 measure, for which the values in the case of the SPARSE group are around zero. Nonetheless, the only significant difference between groups is regarding A2 at electrode F8, in the frontal area.

		\subsubsection{Within session}
\begin{figure}[h!]
	\begin{center}
		\includegraphics[width = 1\linewidth, keepaspectratio]{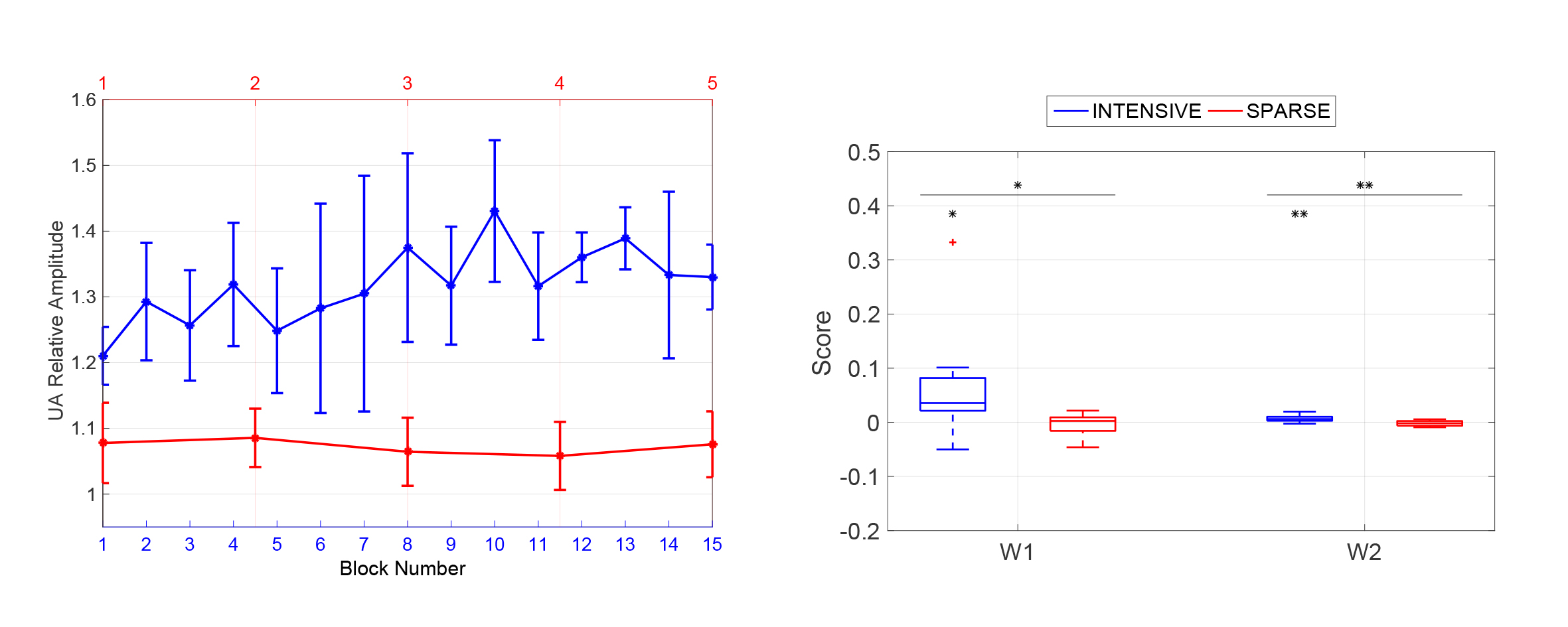}
		\end{center}
		\caption[Evolution within session]{Left: Evolution within session for the INTENSIVE (blue) and SPARSE (red) groups during neurofeedback at Fz;  mean of UA amplitude in all corresponding blocks considering the median of the participants; errors bars show standard deviation. Right: Boxplot with distribution of W1 and W2 measures for the INTENSIVE (blue) and SPARSE (red) groups; the + represents outliers; the * represents p-values below 0.05 and the ** represents p-values below 0.01}
		\label{fig:nfwithin_comparison}
\end{figure}

			\paragraph{Training location}\hfill\\
			Despite the fact that the within-session performance might change along sessions, we still analyzed them globally for all sessions due to their similar progressions over blocks. Therefore, Figure \ref{fig:nfwithin_comparison} refers to one representative session, in which the points represent the mean value for a certain block across sessions, considering the median for all subjects. While there is no visible trend for the SPARSE group, the amplitude tends to increase along blocks for the INTENSIVE group, although it slightly decreases in the last two blocks. 

The distribution of the learning measures may be visualized in Figure \ref{fig:nfwithin_comparison}.  Measures regarding changes within sessions are not significantly different from zero for any of the frequency bands of the SPARSE group, as presented in Table \ref{tab:within_within_all}. The medians of both W1 and W2 for UA were considered significantly larger than zero for the INTENSIVE group, which may be observed both in this figure and in Table \ref{tab:within_within_all}. Moreover this group presents significant differences for both measures when compared with SPARSE group ($U = 68,\ p = 0.014$ and $U = 72,\ p = 0.004$, respectively), visible in Table \ref{tab:within_between_all}. However, for the INTENSIVE group,  W1 and W2 are also significantly different from zero for LA1, IAB and Beta2, and W2 only for LA2. These changes within the INTENSIVE group are also visible when comparing both groups, with significant differences regarding LA1, IAB and Beta2. Therefore, although only the INTENSIVE group reveals changes within session, these are not restricted to the target band.  

\begin{figure}[h!]
	\begin{center}
		\includegraphics[width = 1\linewidth, keepaspectratio]{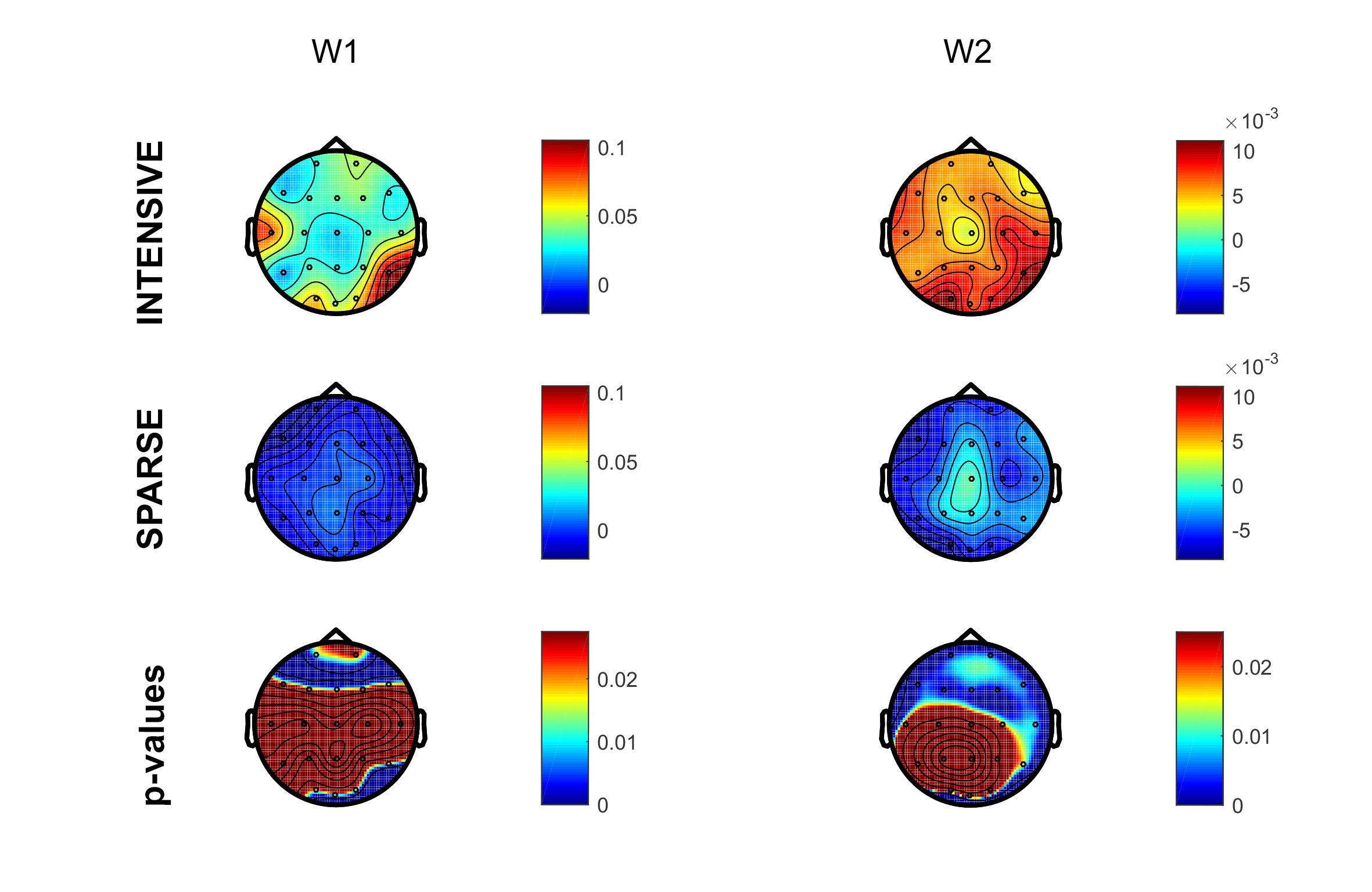}
	\end{center}
	\caption{Topographic distribution of median measures within session for the UA band}
	\label{fig:topoUA_W}
\end{figure}

			\paragraph{Topographic distribution}\hfill\\
Comparing the topographic distributions, in Figure \ref{fig:topoUA_W}, it is visible that, for both measures, the values are higher for the INTENSIVE group, specially in the occipital, temporal and parietal areas, with significant differences between groups. Although not directly targeted by the training, an increase of UA amplitude in these areas might be more easily promoted as the alpha waves occur predominantly in the occipital area.  Nevertheless, as expected due to training, the significant differences between groups occur mainly in the frontal area.

\subsubsection{Learners}
When evaluating only the learners, both A1 and A2 for the UA band have much larger medians than non-learners for both groups, as it was expected considering the rationale used to make this distinction. UA amplitude showed significantly increased performance across sessions in both INTENSIVE (A1: $W = 15,\ p = 0.031$; A2:  $W = 15,\ p = 0.031$) and SPARSE groups (A1: $W = 21,\ p = 0.016$; A2:  $W = 21,\ p = 0.016$). Between groups, A2 regarding UA differs significantly ($U = 28,\ p = 0.017$), with the median being larger in the case of the INTENSIVE group. 

Within sessions, only W2 for UA amplitude in the INTENSIVE group is significantly larger than zero ($W = 15$). This value is significantly different between groups ($U = 30,\ p = 0.040$). Hence, even when considering only the subjects with a positive trend along time, the effects within sessions are exclusive to the INTENSIVE group and along sessions this group reveals stronger changes.

\subsubsection{Transfer trials}
In order to assess if there were a relationship between training performance and the ability to achieve the goal without feedback, the correlation between the A2 measure of training performance during neurofeedback training and transfer trials was computed. For each participant, only the best transfer trial was chosen, which consisted in the one for which the UA amplitude was the largest. Using the Pearson's correlation coefficient, no significant correlations were found (INTENSIVE: $r = 0.278,\ p = 0.469$, SPARSE: $r = -0.103,\ p = 0.793$).  No significant differences were found between groups regarding the best trials ($U = 49,\ p = 0.475$). Thus, it seems that the participants' ability to control their activity outside the training environment was independent of the performance during training. 

\subsubsection{Follow-up} 
Three subjects had to be excluded from follow-up, two due to lack of availability and one due to technical issues during the follow-up session. Thus, the follow-up assessment was performed only on 15 subjects, 8 of the INTENSIVE group and 7 of the SPARSE group. Concerning the UA relative amplitude, regarding the NF period, for both groups most of the participants had lower values during follow-up than during the last session. This is more pronounced in the INTENSIVE group, for which the median is lower and the dispersion to lower negative values is larger. However, the differences between the two time points were not significantly different from zero for any of the groups (INTENSIVE: $W = 6,\ p = 0.109$ and SPARSE: $W = 19,\ p = 0.469$). The results are also not considered significantly different between groups ($U = 2,\ p = 0.054$), yet the obtained p-value is close to 0.05, therefore this outcome should be taken into account. 

\subsection{Baselines}
\begin{figure}[h!]
	\begin{center}
		\includegraphics[width = 1\linewidth, keepaspectratio]{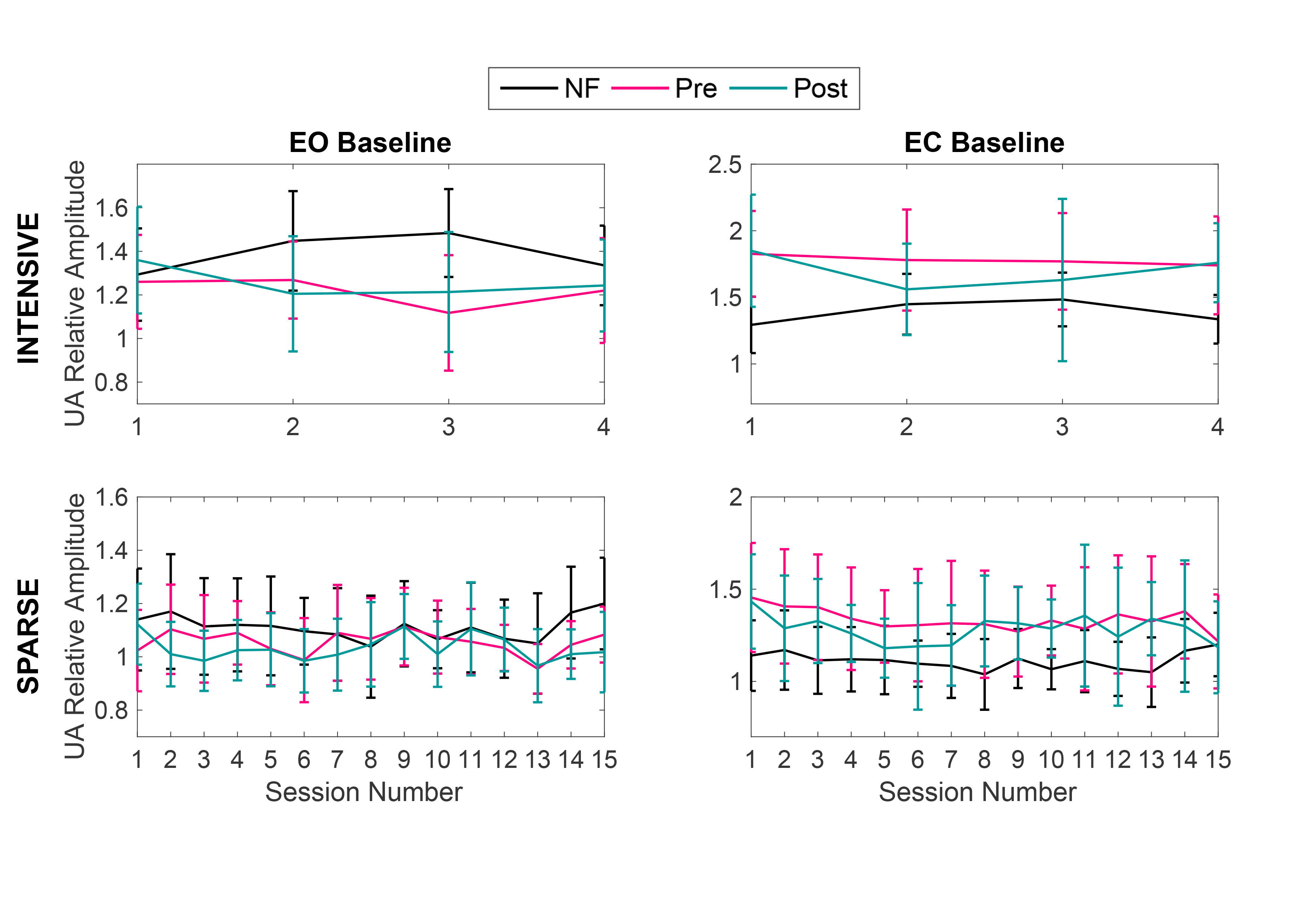}
	\end{center}
		\caption{Comparison of relative amplitudes during pre-NF baseline, post-NF baseline, with EO and EC, and NF (using UA relative amplitudes that are not normalized with the baseline) for INTENSIVE (top) and SPARSE (bottom) groups; error bars show the median absolute deviation}
		\label{fig:baseline}
\end{figure}

	In the data discussed in the present subsection and shown in Figure \ref{fig:baseline}, the NF amplitudes were not normalized, to allow a comparison between the values obtained during training and baselines. While for the INTENSIVE group the median for EO post session was larger than before the session for 3 of the 4 sessions, there is not a marked tendency for the SPARSE group. Moreover, there is no clear trend along time either during EO or EC, and for any of the groups. 
	
	The UA amplitude is predominantly larger during NF than during EO baselines, both pre and post NF. Nonetheless, the values rarely surpass the ones obtained during EC. When evaluating the differences within session, \textit{i.e.} post-baseline $-$ pre-baseline, no p-values under 0.05 were obtained for any participants, both for EO and EC in the two groups. Therefore there is not enough evidence to conclude about significant effects of training on the baselines within session. 
	
The slope of the trendline across sessions was analysed to determine if there was a positive evolution and also to check differences between groups.  No p-values under 0.05 were obtained both for EO and EC, either for the within group or the between groups analysis.

The correlation between pre-baseline values of the first session, prior to any neurofeedback training, and A2 measure was computed to test if this could be used as a predictor of training performance. No significant correlation was found for EO ($r = 0.157,\ p = 0.534$) or EC ($r = 0.105,\ p = 0.678$).

\subsection{Cognitive tests}
Medians of change ($t_f-t_i$) for both groups were not considered significantly different from each other for any of the cognitive tests. However, within group there were score changes whose median was considered significantly larger than zero, namely regarding the Digit Span for both groups and the N-Back with letters only for the SPARSE group. 

The correlation between the performance in cognitive tests and training performance was assessed in order to understand if the results could be seen as NF training effects. Although the subjects in the SPARSE had more chances to practice, the period between assessments was also longer, and there was no evidence that this favoured their performance. A significant correlation was found between the score change in Digit Span Test (Reverse) and A2 for the INTENSIVE group ($r = 0.740,\ p = 0.023$). 

Respecting the follow-up, regardless these effects on the EEG, while for the INTENSIVE group the median score and span of the Digit Span (Forward) test increased in the follow-up session, they decreased for the SPARSE group. For the N-Back test, the performance level either slightly increased or remained the same, and the same happened for the Digit Span (Backward) test with exception of the Span for the INTENSIVE group, which decreased.

\section{Discussion}
\subsection{Individual alpha band}
The individual alpha band was used in this experiment instead of a fixed range of values. 
This decision allows for a more personalized approach and thus should provide a more appropriate training. Yet, it may be also a source of error and variability, as the measurement relied on a single baseline (four interleaved minutes of EO and EC condition) and the manual marks were not straightforward for some of the subjects.
Along time, there were no significant changes of the IAF, as in \citet{zoefel2011} and \citet{escolano2011}.  The only case of intra-variability occurred for the HTF of the SPARSE group. Although this could have been an effect caused by training, which should be further studied, the difficulties in manually marking the frequencies might have triggered this variation. 

\subsection{Trainability}
\citet{zoefel2011} defines trainability as the existence of effects due to training in the target frequency band, which for them was also the upper alpha band. Considering the whole training for all participants, there was no evidence of significant changes in the UA across sessions for any of the groups. Across sessions, none of the training intensity modalities revealed to be effective in producing a growing increase of the UA amplitude.

However, making the distinction between learners and non-learners, and thus considering 56\% and 67\% of the participants of the INTENSIVE and SPARSE groups, respectively, the results point to trainability for both groups. The differences between groups indicate that the INTENSIVE group achieved a significantly higher increase of UA amplitude than the SPARSE group. Equivalent results across sessions were also obtained by \citet{zoefel2011} and \citet{escolano2011} for learners (79\% and 60\%, respectively). Both performed 5 sessions with 5 training blocks of 5 minutes each in five consecutive days. The training load in each session was equivalent to the one for SPARSE group, but the training was done intensively which, combined with the results for this study for learners, suggests that sessions on consecutive days may be more suitable for UA increase.
		
Within session, for the INTENSIVE group, W1 and W2 measures were considered significantly larger than zero regarding UA, which means that this group accomplished a good training performance within sessions. The mean for all sessions decreases for the last 2 blocks suggesting that a larger training session would not benefit training. With that being true, taking out one set of blocks could even be advantageous. Nonetheless, it cannot be discarded that the participants could always be more distracted in the last blocks, no matter the total number, as they might be anticipating the end of the session. For the SPARSE group, however, W1 and W2 were not significantly larger than zero nor there were any visible changes within session for the UA band. Yet, another study, also performing 5 trials with 5 minutes each, found a significant positive tendency for UA (using a metric similar to W2), though only one session was performed \citep{escolano2012}, for what it cannot be equitably compared with the present case (15 sessions).
	
These differences between groups were significant, with the INTENSIVE group obtaining a better performance within session. The equivalent training time of the one received by the SPARSE group, for a single session, is 10 blocks. Since the increasing trend for the INTENSIVE group starts at the first blocks and differences between groups are already visible if one considers only these 10 blocks, it may be concluded that the disparity is mainly due to block duration and not to the total number of blocks. Therefore it seems that smaller blocks, in this case 30 seconds, might be more satisfactory for training when compared to larger ones, of 60 seconds. Additionally, larger breaks might also be beneficial, alleviating the burden of a longer session. Nevertheless, in the work of \citet{kober2017} both stroke patients and healthy controls were able to regulate upper alpha activity within session, with each session consisting of 6 feedback runs of 3 minutes. Even when considering only the learners, within session, there are only significant effects of training for the INTENSIVE group and these are considered significantly stronger  than in the SPARSE group.

\subsection{Independence}
The ability to alter the UA band independently of other frequency bands was evaluated only regarding NF training. Considering all subjects, there were no significant changes across sessions for other frequency bands for any of the groups. Notwithstanding, analyzing across sessions for learners only, independence was accomplished for the INTENSIVE group, but not for the SPARSE, though the differences are for LA2 (close to UA) and IAB (which includes UA).  These results are not in line with what was obtained by \citet{zoefel2011} and \citet{escolano2011}, for whom independence was accomplished regarding training. However, their conclusions are based in different computations. 
\citet{zoefel2011} compared the active pre-baseline of the first session with the last training block (on the 5th session), testing the differences regarding only lower alpha (defined as IAF - 3 to IAF - 1 Hz) and lower beta (defined as IAF + 3 to IAF + 5 Hz), for which no significant effects were found. In the case of \citet{escolano2011}, the comparison was made between the pre-active assessment block of the first and the last session, and the same bands were tested, with no significant differences for any of them. Furthermore, they performed less sessions which, combined with the independence in the case of the INTENSIVE group (also with few sessions), may indicate a larger difficulty in maintaining specificity along time.  
 
Analyzing the results within session, for the SPARSE group there are no significant effects on other frequencies, either considering the whole group or only the learners. However, contrary to what was verified along sessions, for the INTENSIVE group there are significant changes for LA1, LA2 and IAB and also for Beta2. Hence, it is considered that UA was not trained independently from other frequency bands within session for this group, though independence was accomplished when taking into account only the learners. It is possible that the increased LA1 in the INTENSIVE group is due to the need to maintain attention for a longer period of time, which could also cause the participants to remain stiller along the session and have lower Beta2 (as this band is more prone to muscle artifacts). 

\subsection{Baselines}
Contrarily to what was obtained by \citet{escolano2011} with passive baselines, for the upper alpha band, the effects of training did not show up on the baselines along time, neither for EO nor for EC. Nonetheless, they considered only the learners and compared the pre-block of the first and last sessions while this study analyzed the slope along time.  In the work performed by \citet{dekker2014}, 15 sessions with 3 blocks of 8 min were performed on consecutive working days with the aim of increasing the power of the alpha band. The upper alpha power during a passive baseline showed a significant increase between session 1 and 10, but decreased afterwards, which was not observed in this study. 

Considering active baselines, both trainability and independence were observed by \citet{escolano2011} and by \citet{zoefel2011} (where only active baselines were recorded). Likewise, \citet{escolano2012} found significant differences between the NF and the control group concerning active screenings performed before and after a single session. Yet, no significant effects were found within group when comparing the pre-baseline of the training day and the assessment made on the day after. \citet{navarrogil2018} also found significant UA power enhancement during task-related activity comparing the initial and final EEG screenings after 6 sessions (5 trials with 4 minutes each), with 2 sessions per week. Despite this, no effects were observed regarding the EC resting state. 

Regarding clinical applications, \citet{escolano2014a} performed 18 sessions over a month, each with 5 trials of 4 minutes, targeting attention-deficit hyperactivity disorder, and the observed effects were exclusively for task-related activity baseline (EO), they did not show upon resting state (EC). \citet{lavy2018} completed 10 sessions of 30 minutes (10 trials of 3 minutes each) on mild cognitive impaired subjects in order to increase the individual upper alpha. Although there were changes in the peak alpha frequency and memory scores, there was not a significant change of UA power following training.

These results suggest that, regardless of the intensitiy modality, training effects may not appear on resting state baselines. In this case, only a passive baseline was recorded and perhaps the introduction of an active baseline would allow for a better evaluation of the effect of the need to pay attention to a task, which comes alongside with NF training. However, since the active tasks visually resemble the NF training \citep{zoefel2011, escolano2011, escolano2012, escolano2014a, navarrogil2018} it cannot be discarded that, after the participant has experienced training, visualizing feedback elements may trigger changes in activity even if the training is not being applied, and it is considered that this issue should be further investigated.

No predictive value was obtained from the EO and EC regarding the performance during training for UA enhancement, contrarily to what was expected following the work of \citet{wan2014}. They discovered a significant correlation between resting alpha amplitude and learning indices (greatest correlation with the slope of the logarithmic regression along time of the alpha amplitude during sessions). Notwithstanding, the current work used a linear regression instead of logarithmic and was specific to a sub-range of the alpha band. Besides, the fact that the training schedule is also different may as well contribute to this discrepancy.  
 
\subsection{Follow-up}
Despite the fact that the results across sessions (for learners only) and within sessions indicate that the intensive training may be more effective, the follow-up outcomes suggest that the learned control may fade more easily. In fact, the (negative) difference between the follow-up and last session is larger for the INTENSIVE group. Since the training is carried out during 4 days only, while for the SPARSE group it lasts for almost 2 months, it is possible the lack of repetition causes the participants to be more prone to forget how to train. For this reason, occasional ``boosting'' sessions  might be useful to maintain the performance level \citep{angelakis2007}. 
It also has to be taken into consideration that the follow-up session did not have the same duration as the regular session, and was in fact much shorter. For that reason, it might not mirror the training and, in this case, since INTENSIVE's trials were shorter it might not have been enough time for the effects to show.

\subsection{Cognitive tests}
No significant differences were found between groups either for the Digit Span or the N-Back test. However, there were significant effects within each group. Therefore, NF training may have caused improvements but training intensity probably did not influence the results directly. This suggests that the differences regarding NF training were not sufficient to translate into performance. Furthermore, the subjects were young and with high level of education, so there might not be enough space to much greater improvements in one of the groups.   

\subsection{Remarks and limitations}
As training intensity encompasses numerous parameters, such as trial duration, trial number and frequency of the sessions, more groups or distinct experiments would be needed in order to test the influence of each of them on training performance. In this work, an effort was made to choose two contrasting protocols and, at the same time, design them in a way that would be suitable for the participants. However, this causes a problem when trying to isolate the effects of a single parameter and understand, for example, whether the INTENSIVE group achieved a better performance within session due to shorter trials or to the larger number of trials. Furthermore, ideally both groups would be subjected to the same total NF training time, but again by virtue of design constraints it was not feasible, as it would either require an overwhelming load for the INTENSIVE group or too few sessions for the SPARSE. 

The follow-up was designed as a summary of the training, containing a single training block and one transfer trial, besides the cognitive tests and health questionnaire. This choice was made in order to lighten the session so that more volunteers would be willing to participate in the follow-up. However, it impaired a direct comparison with the previous sessions. 

The use of the baselines of each session to compute the normalization may introduce bias in the analysis of NF results if there is a significant effect in baselines along time. Although that was not the case in the present experiment, alternative ways should be explored. 

Furthermore, the fact that A2 values are highly concentrated for the whole SPARSE group, and also the contrast when comparing learners and non-learners across or within sessions, point to the conclusion that this might not be the most suitable way of measuring performance. Otherwise, the performance could be assessed by comparing the first trials of the first session with the last trials of the last session, which considers both the evolution along time and along session.   

Though the choice of the training location was related with working memory, it is also suggested that in future studies the training location is changed to more posterior areas. The amplitude of training frequencies in the alpha range may be particularly difficult to increase frontally since naturally they are decreased relative to posterior areas. Actually, UA enhancement has been tested by addressing P3, Pz, P4, O1 and O2 \citep{escolano2011, zoefel2011, escolano2012, navarrogil2018}.

Moreover, this experiment was performed with healthy subjects, and a specific band and training location were used. Choosing a clinical target or different protocols may cause significant changes regarding the suitability of a certain intensity modality. Additionally, a control group, as well as a larger sample size are recommended to mitigate the impact of inter-individual variations and allow for strongly supported conclusions.

\section{Conclusions}
In the present study, an experiment of NF training for the increase of the activity in the UA band was performed, using two different training intensity modalities. Although further investigation is needed to establish if one modality outperforms the other, the  results suggest that better outcomes might be obtained if each session consists in many short trials (in this case 30 seconds) and is performed on consecutive days. Notwithstanding, if this design is adopted, it must be further examined if it allows the effects to persist. Furthermore, this study targeted healthy subjects and the conclusions may not apply in the case of clinical applications. It is considered that the present work introduces new findings in the field, as to the best of the authors knowledge there are no other studies in which two groups with different training intensity modalities were compared or intensity parameters were exhaustively studied. In the future, we encourage the use of carefully detailed reports of training parameters, as this is a crucial step towards thoroughly studying this topic.  

\section*{Conflict of Interest Statement}
The authors declare that the research was conducted in the absence of any commercial or financial relationships that could be construed as a potential conflict of interest.

\section*{Author Contributions}
IE was responsible for designing and performing the experiment, analysing the results and writing the manuscript. WN guided the analysis of the results and participated in the writing of the manuscript. AC also guided the analysis of the results. CA guided the design of the protocol and the experimental acquisitions. FM contributed with expertise to solve technical issues throughout the course of the experiment. AR supervised the experiment and the writing of the manuscript, providing essential feedback. All authors contributed to the general discussion and review of the manuscript.

\section*{Funding}
This work was supported by the Portuguese Science Foundation (FCT) through Project I\&D LARSyS UID/EEA/50009/2013, funded by national funds through FCT/MCTES (PIDDAC), and Grant BL32/2018\_IST-ID.

\section*{Acknowledgments}
Part of this content first appeared in a Master thesis by \citet{esteves2017}, and is archived at Instituto Superior Técnico (University of Lisbon) website.

\bibliographystyle{apalike}%{ieeetr}
\bibliography{References}

%%% TABLES
% Frequency bands
\begin{table}[H]
\centering
\caption[Frequency Bands]{Frequency ranges definitions}
\label{tab:frequency_bands}
%\renewcommand{\arraystretch}{1}
%\resizebox{0.8\linewidth}{!}{
\begin{tabular}{@{}ll@{}}
\toprule
\textbf{Frequency Bands}            & \textbf{Frequency Range}   \\ \midrule
Individual Theta Band (ITB)         & 4 Hz to LTF                \\
{\begin{tabular}[l]{@{}l@{}}Individual Lower Alpha\\ Band 1 (LA1)\end{tabular}} & LTF to LTF + (IAF - LTF)/2 \\
{\begin{tabular}[l]{@{}l@{}}Individual Lower Alpha\\ Band 2 (LA2)\end{tabular}} & LTF + (IAF - LTF)/2 to IAF \\
{\begin{tabular}[l]{@{}l@{}}Individual Upper Alpha\\ Band (UA)\end{tabular}}    & IAF to HTF \\
Individual Alpha Band (IAB)         & LTF to HTF                 \\
Sensorimotor Rhythm (SMR)           & 12 Hz to 15 Hz             \\
Beta Band (Beta)                    & 16 Hz to 20 Hz             \\
Beta Band 2 (Beta2)                 & 20 to 28 Hz                \\ \bottomrule
\end{tabular}
%}
\end{table}

% Across sessions - within group analysis
\begin{table}[H]
\centering
\caption{Within group analysis across sessions: medians (Mdn), W statistic (W) and p-values (p) resulting from Wilcoxon signed rank test (right-tailed for UA band and two-tailed for the other frequency bands); medians with absolute value below 0.001 are shown as zero}
\label{tab:across_within_all}
\renewcommand{\arraystretch}{1.1}
%\resizebox{0.8\textwidth}{!}{
\begin{tabular}{@{}lrrrrrrrrrrrr@{}}
\toprule
\multicolumn{1}{l}{\multirow{3}{*}{\textbf{\begin{tabular}[l]{@{}l@{}}Frequency\\ bands\end{tabular}}}} & \multicolumn{6}{c}{\textbf{INTENSIVE}} & \multicolumn{6}{c}{\textbf{SPARSE}} \\ \cmidrule(l){2-7} \cmidrule(l){8-13} 
\multicolumn{1}{c}{} & \multicolumn{3}{c}{\textbf{A1}} & \multicolumn{3}{c}{\textbf{A2}} & \multicolumn{3}{c}{\textbf{A1}} & \multicolumn{3}{c}{\textbf{A2}} \\ \cmidrule(l){2-4}  \cmidrule(l){5-7}  \cmidrule(l){8-10}  \cmidrule(l){11-13}
\multicolumn{1}{c}{} & \multicolumn{1}{c}{\textbf{Mdn}} & \multicolumn{1}{c}{\textbf{W}} & \multicolumn{1}{c}{\textbf{p}} & \multicolumn{1}{c}{\textbf{Mdn}} & \multicolumn{1}{c}{\textbf{W}} & \multicolumn{1}{c}{\textbf{p}} & \multicolumn{1}{c}{\textbf{Mdn}} & \multicolumn{1}{c}{\textbf{W}} & \multicolumn{1}{c}{\textbf{p}} & \multicolumn{1}{c}{\textbf{Mdn}} & \multicolumn{1}{c}{\textbf{W}} & \multicolumn{1}{c}{\textbf{p}} \\\midrule %\cmidrule(r){1-1}
\textbf{ITB} & 0.028 & 29 & 0.496 & 0 & 21 & 0.910 & 0.028 & 31 & 0.359 & 0.002 & 23 & 1.000 \\
\textbf{LA1} & -0.038 & 18 & 0.652 & -0.019 & 19 & 0.734 & 0.124 & 37 & 0.098 & 0.004 & 34 & 0.203 \\
\textbf{LA2} & 0.034 & 25 & 0.820 & 0.014 & 25 & 0.820 & 0.084 & 33 & 0.250 & 0.002 & 34 & 0.203 \\
\textbf{UA} & 0.011 & 24 & 0.455 & 0.003 & 27 & 0.326 & 0.059 & 33 & 0.125 & 0.001 & 26 & 0.367 \\
\textbf{IAB} & 0.016 & 22 & 1.000 & 0.005 & 25 & 0.820 & 0.075 & 37 & 0.098 & 0.002 & 34 & 0.203 \\
\textbf{SMR} & 0.009 & 34 & 0.203 & 0.010 & 36 & 0.129 & -0.042 & 10 & 0.164 & -0.002 & 13 & 0.301 \\
\textbf{Beta} & -0.028 & 18 & 0.652 & -0.014 & 18 & 0.652 & -0.043 & 16 & 0.496 & 0 & 21 & 0.910 \\
\textbf{Beta2} & -0.002 & 18 & 0.652 & -0.006 & 20 & 0.820 & -0.053 & 11 & 0.203 & -0.001 & 13 & 0.301 \\ \bottomrule
\end{tabular}
%}
\end{table}

%Across sessions - between groups analysis
\begin{table}[H]
\centering
\caption[Between groups analysis across sessions]{Between groups analysis across sessions concerning the learning measures for different frequencies at the training location Fz: U statistic (U) and p-values (p) resulting from Wilcoxon rank sum Test}
\label{tab:across_between_all}
\begin{tabular}{lrrrr}
\toprule
\multirow{2}{*}{\textbf{\begin{tabular}[c]{@{}l@{}}Frequency\\   bands\end{tabular}}} & \multicolumn{2}{c}{\textbf{A1}}         & \multicolumn{2}{c}{\textbf{A2}}         \\ \cmidrule(l){2-3} \cmidrule(l){4-5}
                                                                                      & \multicolumn{1}{c}{\textbf{U}} &  \multicolumn{1}{c}{\textbf{p}} & \multicolumn{1}{c}{\textbf{U}} & \multicolumn{1}{c}{\textbf{p}} \\\midrule%\cmidrule(r){1-1}% \cmidrule(l){2-5}
\textbf{ITB}                                                                          & 36                   & 0.730            & 42                   & 0.931            \\
\textbf{LA1}                                                                          & 21                   & 0.094            & 27                   & 0.258            \\
\textbf{LA2}                                                                          & 29                   & 0.340            & 45                   & 0.730            \\
\textbf{UA}                                                                           & 33                   & 0.546            & 46                   & 0.667            \\
\textbf{IAB}                                                                          & 23                   & 0.136            & 41                   & 1.000            \\
\textbf{SMR}                                                                          & 63                   & 0.050            & 63                   & 0.050            \\
\textbf{Beta}                                                                         & 41                   & 1.000            & 36                   & 0.730            \\
\textbf{Beta2}                                                                        & 46                   & 0.667            & 42                   & 0.931            \\ \bottomrule
\end{tabular}
\end{table}

% Within session - within group analysis		
\begin{table}[H]
\centering
\caption[Within group analysis within session]{Within group analysis within session: median (Mdn), W statistic (W) and p-values (p) resulting from Wilcoxon signed rank test (right-tailed for UA band and two-tailed for the other frequency bands); medians with absolute value below 0.001 are shown as zero}
\label{tab:within_within_all}
%\renewcommand{\arraystretch}{1}
%\resizebox{0.8\textwidth}{!}{
\begin{tabular}{@{}lrrrrrrrrrrrrr@{}}
\toprule
\multirow{3}{*}{\textbf{\begin{tabular}[c]{@{}l@{}}Frequency\\   Bands\end{tabular}}} & \multicolumn{6}{c}{\textbf{INTENSIVE}} & \multicolumn{6}{c}{\textbf{SPARSE}} \\ \cmidrule(r){2-7} \cmidrule(r){8-13}
 & \multicolumn{3}{c}{\textbf{W1}} & \multicolumn{3}{c}{\textbf{W2}} & \multicolumn{3}{c}{\textbf{W1}} & \multicolumn{3}{c}{\textbf{W2}} \\ \cmidrule(l){2-4}  \cmidrule(l){5-7}  \cmidrule(l){8-10}  \cmidrule(l){11-13}
 & \multicolumn{1}{c}{\textbf{Mdn}} & \multicolumn{1}{c}{\textbf{W}} & \multicolumn{1}{c}{\textbf{p}} & \multicolumn{1}{c}{\textbf{Mdn}} & \multicolumn{1}{c}{\textbf{W}} & \multicolumn{1}{c}{\textbf{p}} & \multicolumn{1}{c}{\textbf{Mdn}} & \multicolumn{1}{c}{\textbf{W}} & \multicolumn{1}{c}{\textbf{p}} & \multicolumn{1}{c}{\textbf{Mdn}} & \multicolumn{1}{c}{\textbf{W}} & \multicolumn{1}{c}{\textbf{p}} \\\midrule %\cmidrule(r){1-1}
\textbf{ITB} & 0.003 & 22 & 1.000 & 0 & 19 & 0.734 & -0.001 & 24 & 0.910 & 0 & 22 & 1.000 \\
\textbf{LA1} & 0.113 & 40 & \textbf{0.039} & 0.007 & 44 & \textbf{0.008} & -0.002 & 20 & 0.820 & 0.005 & 28 & 0.570 \\
\textbf{LA2} & 0.058 & 39 & 0.055 & 0.009 & 44 & \textbf{0.008} & -0.001 & 20 & 0.820 & 0.004 & 24 & 0.910 \\
\textbf{UA} & 0.036 & 40 & \textbf{0.020} & 0.006 & 43 & \textbf{0.006} & 0.003 & 17 & 0.752 & -0.002 & 14 & 0.850 \\
\textbf{IAB} & 0.067 & 45 & \textbf{0.004} & 0.008 & 45 & \textbf{0.004} & 0 & 21 & 0.910 & 0.002 & 24 & 0.910 \\
\textbf{SMR} & -0.005 & 21 & 0.910 & 0 & 12 & 0.250 & -0.001 & 20 & 0.820 & -0.001 & 22 & 1.000 \\
\textbf{Beta} & 0.018 & 25 & 0.820 & 0 & 15 & 0.426 & 0.009 & 36 & 0.129 & 0.003 & 38 & 0.074 \\
\textbf{Beta2} & -0.042 & 4 & \textbf{0.027} & -0.005 & 0 & \textbf{0.004} & -0.004 & 14 & 0.359 & -0.001 & 16 & 0.496\\
\bottomrule
\end{tabular}
%}
\end{table}		

% Within session - between groups analysis				
\begin{table}[H]
\centering
\caption[Between groups analysis within session]{Between groups analysis within session concerning the learning measures for different frequencies at the training location Fz:  U statistic (U) and p-values (p) resulting from Wilcoxon rank sum test}
\label{tab:within_between_all}
\begin{tabular}{lrrrr}
\toprule
\multirow{2}{*}{\textbf{\begin{tabular}[c]{@{}l@{}}Frequency\\   bands\end{tabular}}} & \multicolumn{2}{c}{\textbf{W1}}         & \multicolumn{2}{c}{\textbf{W2}}         \\ \cmidrule(l){2-3} \cmidrule(l){4-5}
                                                                                      & \multicolumn{1}{c}{\textbf{U}} & \multicolumn{1}{c}{\textbf{p}} & \multicolumn{1}{c}{\textbf{U}} & \multicolumn{1}{c}{\textbf{p}} \\ \midrule%\cmidrule(r){1-1}
\textbf{ITB}                                                                          & 37                   & 0.796            & 38                   & 0.863            \\
\textbf{LA1}                                                                          & 65                   & \textbf{0.032}   & 45                   & 0.730            \\
\textbf{LA2}                                                                          & 63                   & 0.050            & 55                   & 0.222            \\
\textbf{UA}                                                                           & 68                   & \textbf{0.014}   & 72                   & \textbf{0.004}   \\
\textbf{IAB}                                                                          & 78                   & \textbf{$<$0.001}   & 74                   & \textbf{0.002}   \\
\textbf{SMR}                                                                          & 40                   & 1.000            & 39                   & 0.931            \\
\textbf{Beta}                                                                         & 40                   & 1.000            & 18                   & 0.050            \\
\textbf{Beta2}                                                                        & 17                   & \textbf{0.040}   & 24                   & 0.162    \\
\bottomrule       
\end{tabular}
\end{table}	

%%% FIGURES

% across comparison along time and boxplot

% topoplot across

% within comparison along time

%topoplot within

% baselines EO, EC, NF

% follow-up boxplot
\begin{figure}[h!]
	\begin{center}
		\includegraphics[width = 0.5\linewidth, keepaspectratio]{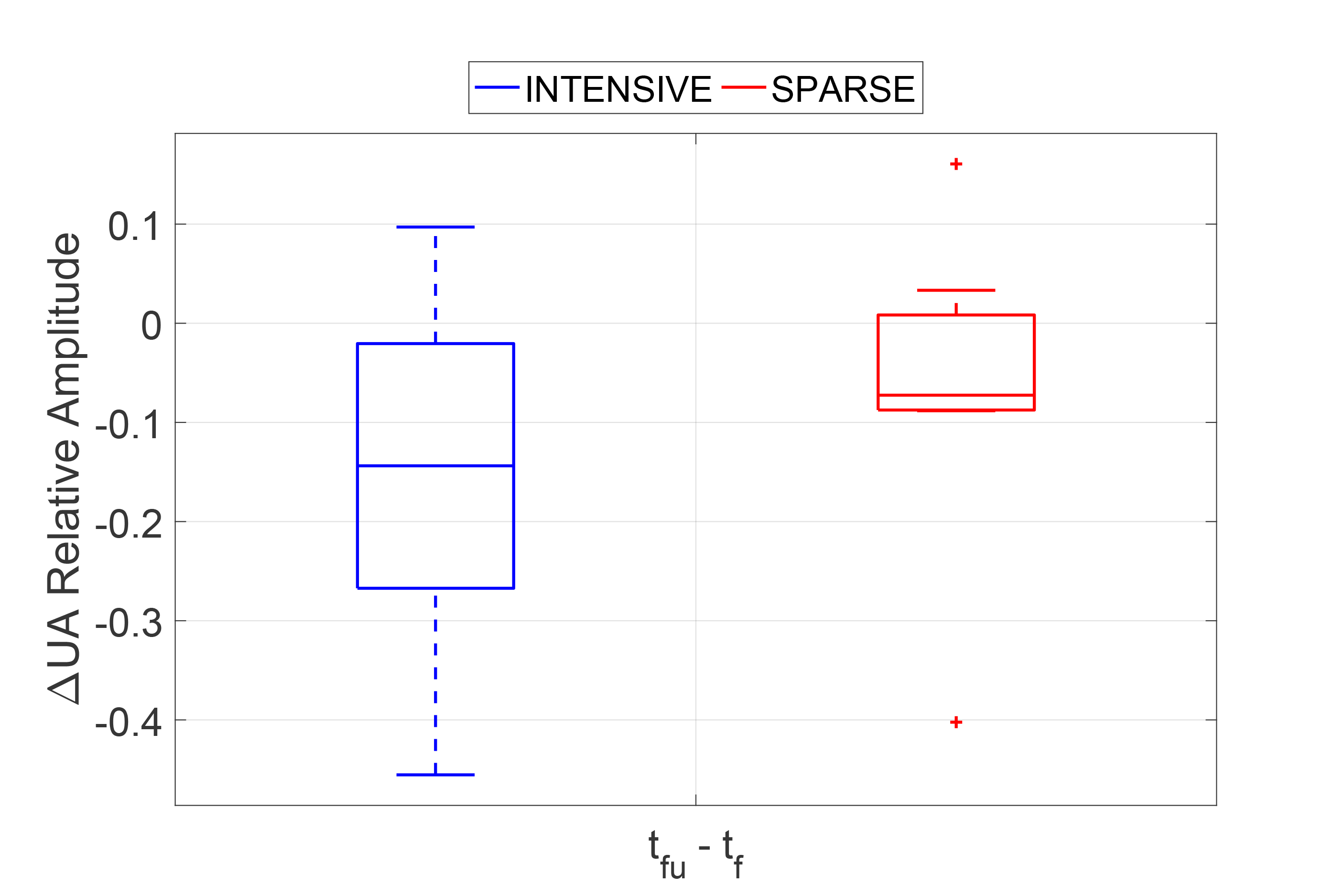}
	\end{center}		
		\caption[Evolution within sessions]{Boxplot with distribution of $t_{fu}-t_f$ UA amplitudes for the INTENSIVE (blue) and SPARSE (red) groups; the + represents outliers}
		\label{fig:fu_tftfuNF}
\end{figure}

\end{document}